\documentstyle[12pt]{article}

\def\agt{\buildrel {\mbox{$>$}} \over {\raisebox{-0.8ex}{\hspace{-0.05in}
$\sim$}}}
\def\alt{\buildrel {\mbox{$<$}} \over {\raisebox{-0.8ex}{\hspace{-0.05in}
$\sim$}}}
\def\overlay#1#2{\ifmmode%
\setbox0=\hbox{$#1$}%
\setbox1=\hbox to\wd0{\hss$#2$\hss}\else%
\setbox0=\hbox{#1}%
\setbox1=\hbox to\wd0{\hss#2\hss}\fi%
 #1\hskip-\wd0\box1 }

\catcode`@=11
\newcount\@tempcntc
\def\@citex[#1]#2{\if@filesw\immediate\write\@auxout{\string\citation{#2}}\fi
  \@tempcnta\z@\@tempcntb\m@ne\def\@citea{}\@cite{\@for\@citeb:=#2\do
    {\@ifundefined
       {b@\@citeb}{\@citeo\@tempcntb\m@ne\@citea\def\@citea{,}{\bf ?}\@warning
       {Citation `\@citeb' on page \thepage \space undefined}}%
    {\setbox\z@\hbox{\global\@tempcntc0\csname b@\@citeb\endcsname\relax}%
     \ifnum\@tempcntc=\z@ \@citeo\@tempcntb\m@ne
       \@citea\def\@citea{,}\hbox{\csname b@\@citeb\endcsname}%
     \else
      \advance\@tempcntb\@ne
      \ifnum\@tempcntb=\@tempcntc
      \else\advance\@tempcntb\m@ne\@citeo
      \@tempcnta\@tempcntc\@tempcntb\@tempcntc\fi\fi}}\@citeo}{#1}}
\def\@citeo{\ifnum\@tempcnta>\@tempcntb\else\@citea\def\@citea{,}%
  \ifnum\@tempcnta=\@tempcntb\the\@tempcnta\else
   {\advance\@tempcnta\@ne\ifnum\@tempcnta=\@tempcntb \else \def\@citea{--}\fi
    \advance\@tempcnta\m@ne\the\@tempcnta\@citea\the\@tempcntb}\fi\fi}
\catcode`@=12
\begin{document}

\hfill
{\vbox{
\hbox{CPP-94-37}
\hbox{DOE-ER-40757-060}
\hbox{UCD-95-4}
\hbox{February 1995}}}

\begin{center}
{\large \bf Hadronic Production of S-wave and P-wave  \\
Charmed Beauty Mesons via Heavy Quark Fragmentation}

\vspace{0.1in}

Kingman Cheung\footnote{Internet address: {\tt cheung@utpapa.ph.utexas.edu}}

{\it Center for Particle Physics, University of Texas at Austin,
Austin TX 78712 U.S.A.}

Tzu Chiang Yuan \footnote{Internet address: {\tt yuantc@ucdhep.ucdavis.edu}}

{\it Davis Institute for High Energy Physics, Department of Physics, \\
University of California at Davis, Davis CA 95616 U.S.A.}

\end{center}

\begin{abstract}

At hadron colliders the dominant production mechanism of $(\bar bc)$  mesons
with large transverse momentum is due to parton fragmentation.
We compute  the rates and transverse
momentum spectra for production of S-wave and P-wave $(\bar b c)$ mesons
at the Tevatron via the direct fragmentation of
the bottom antiquark as well as the Altarelli-Parisi induced
gluon fragmentation. Since all the radially and orbitally excited
$(\bar b c)$ mesons below the $BD$ flavor threshold will cascade
into the pseudoscalar ground state $B_c$ through
electromagnetic and/or hadronic transitions, they all contribute to
the inclusive production of $B_c$.
The contributions of the excited S-wave and P-wave states
to the inclusive production of $B_c$ are 58 and 23\%,
respectively, and hence significant.

\end{abstract}

\thispagestyle{empty}

\newpage

\section{Introduction}

Bound states composed of two heavy quarks such as
$(c\bar c),\, (b\bar b)$ and $(\bar b c)$ mesons,
and $(ccq),\, (bbq)$ and $(bcq)$ baryons ($q$ denotes a light quark)
are of interests because they can
be produced with sizable rates at the current high energy hadron
or $e^+e^-$ colliders.  The Standard Model predictions of the production
rates of these bound states
can therefore be confronted with experimental data.

The direct production of heavy mesons like heavy quarkonia and $(\bar b c)$
bound states can provide very interesting tests for
perturbative QCD.  The production of $J/\psi$ and $\psi'$ \cite{CDF}
at the Tevatron has already raised a lot of theoretical interests in
explaining the excess of the experimental data above the lowest order
perturbative QCD  calculation \cite{stirling}, especially
at the large transverse momentum $(p_T)$ region. The ideas of heavy quark and
gluon fragmentation \cite{gswave,charm,bcfrags,gpwave,qpwave,bcfrags_others}
have been successfully applied to explain the experimental data
of the prompt $J/\psi$ production from CDF within
a factor of five \cite{BDFM,CG,RS,CWT}.
Various attempts \cite{close,chowise,psipi,psipi2}
using the same fragmentation ideas
have also been made to resolve the
$\psi^\prime$ surplus problem observed at CDF.

Recently, the preliminary CDF results \cite{papadimi} also showed that
the production rates of the $1S$, $2S$, and $3S$ $\Upsilon$ states
are in excess of the leading order calculation.
While the $\Upsilon(1S)$ and $\Upsilon(2S)$ results can
be partly explained by including the fragmentation contribution,
the $\Upsilon(3S)$ result showed an excess of about an order of magnitude
over the QCD prediction even with
the fragmentation contribution included \cite{papadimi}.
One subtlety is that the relevant $p_T$ for the fragmentation
contribution to dominate should be larger in the bottomonium system than in
the charmonium system, such that fragmentation is only valid for
$p_T \agt \,$(1--2)$\, m_b$ in the former case.
Besides, one has to worry about the very small $p_T$ region because,
unlike the charmonium system,
the experimental triggering conditions on the muon pair coming from the
bottomonium leptonic decay allow  experimentalists to
measure the transverse momentum of the
bottomonium all the way down to about 0--1 GeV \cite{papadimi}.
In order to fully understand the $p_T$ spectrum of the bottomonium production,
different production mechanisms have to be brought into picture to
explain the production rates in different $p_T$ regions.

The $(\bar b c)$ meson system, which is {\it intermediate} between the
$J/\psi$ and $\Upsilon$ families, is also an interesting physical system
to study. The mass spectrum of the $(\bar b c)$ mesons
can be predicted reliably from quarkonium potential models
\cite{eitqui,baganetal,kiselevetal} without introducing any new parameters
and their  decay constants can be computed using QCD spectrum sum
rules \cite{baganetal,kiselevetal}.
We will adopt the Particle Data Group \cite{PDG} conventions,
denoting the $1S$ pseudoscalar ($^1S_0$) and vector meson ($^3S_1$)
$(\bar b c)$ states by $B_c$ and $B_c^*$, respectively.
Higher radially and orbitally  excited states are labeled by
the standard spectroscopy notation: $n\,^{2S+1}L_J$, where the integer $n$
is the principal quantum number, and $L$, $S$, and $J=L+S$  are respectively
the orbital angular momentum, total spin, and total angular momentum of the
bound state.
In the $LS$ coupling scheme, for each principal quantum number $n$,
the spin-singlet and the spin-triplet S-wave ($L=0$) states are denoted by
$^1S_0$ and $^3S_1$, respectively.  Those for the P-wave ($L=1$)
and D-wave ($L=2$) states are denoted by $^1P_1$, $^3P_J$ ($J=0,1,2$),
and $^1D_2$, $^3D_J$ ($J=1,2,3)$, respectively.

According to the results of the potential model calculation \cite{eitqui},
the first two sets ($n=1$ and $n=2$) of S-wave states, the first ($n=1$)
and probably the entire second set ($n=2$) of P-wave states, and
the first set ($n=1$)  of D-wave states lie below the $BD$ flavor
threshold.  Since QCD interactions are diagonal in flavors,
the annihilation channel of excited $(\bar b c)$ mesons can only occur
through the weak gauge boson ($W$) exchange and  is therefore suppressed
relative to the electromagnetic and hadronic transitions to other
lower lying states.
The excited states below the $BD$ threshold  will cascade down into
the ground state $B_c$ via emission of photons and/or pions, while the
other states above the $BD$ threshold will decay rapidly into a pair of
$B$ and $D$ mesons. Inclusive production of the
$B_c$ meson therefore includes the production of the
$n=1$ and $n=2$ S-wave and P-wave states, and the $n=1$ D-wave
states.

The production of the S-wave ($^1S_0$ and $^3S_1$)
states were first computed exactly to leading order in Ref.~\cite{bc_ee}
at the $e^+e^-$ machine, in particular at the $Z$ resonance.
Later, it was realized \cite{bcfrags,bcfrags_others}
that the complicated formulas in these complete calculations
can be simplified by a factorization
approach. The dominant contribution in the leading order calculation can be
factorized into a short distance piece,
which describes the partonic process of the decay of $Z$ into a high energy
$b \bar b$ pair, and a fragmentation piece
describing how the $\bar b$ antiquark splits into the two S-wave states. The
corresponding  fragmentation functions
$D_{\bar b \to B_c}(z)$ and  $D_{\bar b \to B_c^*}(z)$, which are
independent of the short-distance piece,
were shown to be calculable by  perturbative QCD at the heavy quark mass scale
\cite{bcfrags}.  Recently, the production of the S-wave states has also
been computed at hadronic colliders like the Tevatron and the
Large Hadron Collider (LHC) both by a complete ${\cal O}(\alpha_s^4)$
calculation \cite{changchen,slab,bere,mase,klr} and by using the simpler
fragmentation  approach \cite{cheung,induceglue}. We note
that unlike the $J/\psi$ production in hadronic collisions
in which the major contributions come from
(1) $g\to \chi_{cJ}$ fragmentation followed by the decay
$\chi_{cJ} \to J/\psi + \gamma$ \cite{BDFM,CG,RS,CWT}, and
(2) gluon fragmentation into a color-octet $(c \bar c)$ $^3S_1$ state which
subsequently evolves nonperturbatively into $J/\psi$ \cite{psipi,psipi2},
the fragmentation diagrams for producing
$(\bar b c)$ mesons form a subset of the
whole set of ${\cal O}(\alpha_s^4)$  diagrams, which are the
{\it leading-order} diagrams for producing $(\bar b c)$ mesons
in hadronic collisions.
It is therefore not clear that if the production of $(\bar b c)$ mesons
in hadronic collisions is dominated by parton fragmentation.
However, detailed calculations by  Chang {\it et al.} \cite{changchen},
Slabospitsky \cite{slab}, and Kolodziej {\it et al} \cite{klr}
indicated that the fragmentation approach is valid for the S-wave
production at the large transverse momentum region.
We will discuss more about this later in the closing section.

In this paper, we study the hadronic
production of  $(\bar bc)$ mesons.
We compute the  production rates of the S-wave and
P-wave $(\bar b c)$ mesons at the Tevatron using the fragmentation
approach. Intuitively, the dominant production mechanism
of the $(\bar b c)$ mesons at the large transverse momentum region
must arise from the direct fragmentation of the heavy $\bar b$ antiquark.
The relevant question is whether experiments can probe
the transverse momentum region  where fragmentation dominates.
Unfortunately, to answer this question it also requires a complete
${\cal O}(\alpha_s^4)$  calculation for the production of the
P-wave states, which is not available at the moment.
Here we  assume that the fragmentation approach also works for production of
the P-wave $(\bar b c)$ mesons in the transverse momentum range that
is being probed experimentally at the Tevatron.
In this work, we do not include the contributions from the D-wave
states because the corresponding fragmentation functions
are very small \cite{dwave}.
Although the  production  of the D-wave states
is of great interest by themselves, they  only
contribute about 2 \% \cite{dwave} to the inclusive production of
$B_c$.
In the near future, like other heavy quarkonia,
the production of $(\bar bc)$ mesons at the Tevatron
may therefore provide another interesting test for perturbative QCD.
Although we will only show our results for the positively charged states
$(\bar b c)$, all the results presented in this work also hold
for the negatively charged states $(b \bar c)$.

The organization of this paper is as follows.
In the next section, we discuss in detail the general procedures
to calculate the  production cross sections using the fragmentation approach.
In Sec. III we present the transverse momentum spectra and the
integrated  cross sections
for the production of S-wave and P-wave $(\bar b c)$
states, as well as  the inclusive production rate for the $B_c$ meson.
Discussions and conclusions are made in Sec. IV.
For completeness  we also collect all the S-wave and
P-wave fragmentation functions at the heavy-quark mass scale in the
Appendix.

Before leaving this preamble, we note that
a preliminary result from CDF  had provided
a hint for the $B_c$ existence by looking at the production rate of
$J/\psi+\pi$ in the mass bin of 6.1--6.4 GeV \cite{private}.

\section{Inclusive production cross sections in the fragmentation approach}

Theoretical calculations of production cross sections in high
energy hadronic collisions are based on the idea of
factorization.
Factorization divides an inclusive or exclusive hadronic production process
into short-distance pieces and long-distance pieces.
The short-distance pieces are
perturbatively calculable to any desired accuracy in QCD, while
the long-distance pieces are in general not calculable
within perturbation theory but can be parameterized as phenomenological
functions, which can be determined by experiments.
The factorization used here for the
production of $(\bar b c)$ mesons divides the process into the
production of a high energy parton (a $\bar b$ antiquark or a gluon) and
the fragmentation of this parton into various $(\bar b c)$ states.
The novel feature in our approach, which is due to a recent
theoretical development \cite{bcfrags,qpwave}, is
that the relevant fragmentation
functions at the heavy quark mass scale can be calculated
in perturbative QCD to any desired accuracy.
This is easily understood from the fact that the fragmentation of a
$\bar b$ antiquark into a $(\bar b c)$ meson involves
the creation of a $c \bar c$ pair out of the vacuum.
The natural scale for this particular hadronization is of
order of the mass of the quark pair being created.
In our case, this scale is of order $m_c$, which is considerably larger than
$\Lambda_{\rm QCD}$. One can therefore calculate reliably
the fragmentation function as an series expansion in the strong coupling
constant $\alpha_s$ using perturbative QCD.

The production of  high energy partons  also involves the factorization
into  the parton distribution functions inside the hadrons and
the parton-parton hard scattering.
Let $H$ denotes any $(\bar b c)$ meson states.   The differential
cross section $d\sigma/dp_T$ versus the transverse momentum $p_T$ of $H$
is given by
\begin{eqnarray}
\frac{d \sigma}{d p_T}(p\bar p \to  H(p_T) X) & = & \sum_{ij}
\int dx_1 dx_2 dz f_{i/p}(x_1,\mu) f_{j/\bar p}(x_2,\mu)
\left [
\frac{d \hat \sigma}{dp_T} (ij \to \bar b(p_T/z)X,\,\mu) \nonumber \right. \\
&& \left. \times
D_{\bar b \to H} (z,\mu)
+ \frac{d\hat \sigma}{dp_T} (ij\to g(p_T/z)X,\mu)\; D_{g\to H}
(z,\mu) \right ] \; .
\label{*}
\end{eqnarray}
The physical interpretation is as follows: a heavy $\bar b$ antiquark or
a gluon is produced in a hard process with a transverse momentum $p_T/z$
and then it fragments into $H$ carrying a longitudinal momentum fraction
$z$.   We assume that $H$ is moving in the same direction as the fragmenting
parton. In the above equation, $f_{i/p(\bar p)}(x,\mu)$'s
are the parton distribution functions,  $d \hat \sigma$'s
represent the subprocess cross sections, and $D_{i\to H}(z,\mu)$'s
are the parton fragmentation functions at the scale $\mu$.
For production of the $\bar b$ antiquark,
we include the subprocesses $gg\to b\bar b$, $g \bar b \to g\bar b$,
and $q \bar q \to b\bar b$; while for the production of the gluon $g$, we
include the subprocesses $gg\to gg$, $q\bar q\to gg$, and
$g q(\bar q)\to g q(\bar q)$.
In Eq.~(\ref{*}), the factorization scale $\mu$ occurs in the parton
distribution functions, the subprocess cross sections, and the
fragmentation functions.  In general, we can choose three different scales
for these three entities.  For simplicity and ease of estimating  the
uncertainties due to changes in scale, we choose a common scale $\mu$
for  all three of them.

The physical production rates should be independent of choices of the
scale $\mu$, because $\mu$ is just an artificial entity introduced
to factorize the whole process into different parts in the renormalization
procedure. However, this independence of
scale can only be achieved if both the production of the high energy
partons and the fragmentation functions are calculated to all orders in
$\alpha_s$.   So far, only the next-to-leading order $\hat \sigma$'s and
the leading order fragmentation functions are available, so the
production cross sections do depend on the choice of $\mu$ to a certain
degree.  We will estimate the dependence on $\mu$ by varying the scale
$\mu=(0.5-2)\mu_R$, where $\mu_R$ is our primary choice of scale
\begin{equation}
\label{scale}
\mu_R = \sqrt{p_T^2({\rm parton}) + m^2_b} \;.
\end{equation}
This choice of scale, which is of order $p_T({\rm parton})$, avoids the
large logarithms in the short-distance part $\hat \sigma$'s.  However,
we have to sum up the logarithms of order $\mu_R/m_b$ in the
fragmentation functions.  But this can be implemented by evolving the
Altarelli-Parisi equations for the fragmentation functions.

\subsection{Evolution of Fragmentation Functions}

The Altarelli-Parisi
evolution equations for the fragmentation functions are
\begin{equation}
\label{Db}
\mu \frac{\partial}{\partial \mu} D_{\bar b\to H}(z,\mu) =
\int_z^1 \frac{dy}{y}
P_{\bar b\to \bar b}(z/y,\mu)\; D_{\bar b \to H}(y,\mu) +
\int_z^1 \frac{dy}{y} P_{\bar b\to g}(z/y,\mu)\; D_{g \to H}(y,\mu) \,,
\end{equation}
\begin{equation}
\label{Dg}
\mu \frac{\partial}{\partial \mu} D_{g\to H}(z,\mu) = \int_z^1 \frac{dy}{y}
P_{g \to \bar b}(z/y,\mu)\; D_{\bar b \to H}(y,\mu) +
\int_z^1 \frac{dy}{y} P_{g \to g}(z/y,\mu)\; D_{g \to H}(y,\mu) \,,
\end{equation}
where $H$ denotes any $(\bar b c)$ states, and $P_{i\to j}$ are the usual
Altarelli-Parisi splitting functions.
The leading order expressions for $P_{i\to j}$ can be found in the Appendix.

The boundary conditions for solving the above Altarelli-Parisi
equations are the fragmentation
functions $D_{\bar b \to H}(z,\mu_0)$ and $D_{g\to H}(z,\mu_0)$ that we
can calculate by perturbative QCD at the initial scale $\mu_0$,
which is of the order of the $b$-quark mass.
At present, all the S-wave \cite{bcfrags} and P-wave \cite{qpwave}
fragmentation functions for $\bar b\to (\bar b c)$ have been calculated
to leading order in $\alpha_s$.  They are all collected in the Appendix.
The initial scale $\mu_0$ for $D_{\bar b \to H}(z,\mu_0)$
is chosen to be $\mu_0=m_b + 2m_c$, which is the minimum virtuality of
the fragmenting $\bar b$ antiquark \cite{bcfrags}.
On the other hand, since the initial gluon fragmentation function
$D_{g\to H}(z,\mu_0)$ is suppressed by one extra power of $\alpha_s$
relative to $D_{\bar b\to H}(z,\mu_0)$, we simply choose the initial gluon
fragmentation function to be $D_{g\to H}(z,\mu_0)=0$
for $\mu_0 \le 2(m_b+m_c)$ --- the minimum virtuality
of the fragmenting gluon \cite{induceglue}.
We can also examine the relative importance of these fragmentation functions.
The initial $D_{\bar b\to H}(z,\mu_0)$ is of order $\alpha_s^2$, whereas
the initial $D_{g\to H}(z,\mu_0)$
is of order $\alpha_s^3$ and has been set to be zero for
$\mu_0 \leq 2(m_b + m_c)$ as discussed above.
But when the scale is evolved up to  a higher scale $\mu$,
$D_{\bar b\to H}(z,\mu)$ is still of order $\alpha_s^2$, while the
induced $D_{g\to H}(z,\mu)$ is of order $\alpha_s^3 \log(\mu/\mu_0)$.
At a sufficiently large scale $\mu$ the logarithmic enhancement can
offset the extra suppression factor of $\alpha_s$. Thus the
Altarelli-Parisi induced gluon fragmentation functions can be as
important as the $\bar b$ antiquark fragmentation, even though
the initial gluon fragmentation functions are suppressed. While these
Altarelli-Parisi induced gluon fragmentation functions play only
a moderate role at the Tevatron, they will play a more significant
role at the LHC \cite{induceglue}.

To obtain the fragmentation functions at an arbitrary scale greater than
$\mu_0$, we numerically integrate the Altarelli-Parisi evolution equations
(\ref{Db})--(\ref{Dg})  with the boundary conditions described in the
above paragraph.
Since the initial light-quark fragmentation
functions $D_{q \to H}(z,\mu_0)$ are of order $\alpha_s^4$, one can set them
to be zero as well for $\mu_0 \leq 2(m_b + m_c)$.
One may ask if the light-quark fragmentation functions
can be induced in the same manner as in the gluon case by Altarelli-Parisi
evolution.   Since both the Altarelli-Parisi splitting functions
$P_{q\to q}$ and $P_{q\to g}$ are total plus-functions, the induced
light-quark fragmentation functions $D_{q \to H}(z)$ can only be
total plus-functions or vanishing identically. We do not
anticipate that these induced light-quark fragmentation functions,
if nontrivial, will play any significant role in our analysis.

\section{Numerical results}

Leading order QCD formulas are employed for the parton-level scattering
cross sections and CTEQ(2M) \cite{cteq} is used
for the initial parton distributions.
The inputs to the initial fragmentation functions are  the heavy quark masses
$m_b$ and $m_c$, and the nonperturbative parameters associated  with
the wavefunctions of the bound states.
For the S-wave states there is only one nonperturbative parameter,
which is the radial wavefunction $R_{nS}(0)$ at the origin.  However for
the P-wave fragmentation functions we have two nonperturbative parameters
$H_1$ and $H_8^\prime$ associated with the color-singlet and
the color-octet mechanisms, respectively \cite{bbl}.
Two of the P-wave states ($^1P_1$ and $^3P_1$) are mixed to
form the two physical states, denoted by $|1^+ \rangle$
and $|1^{+\prime}\rangle$.  Further details of the mixings can be found in
Refs.~\cite{qpwave,eitqui}.  The above input parameters for the fragmentation
functions are summarized in Table~\ref{table1}.

As mentioned in the previous section, we have set the scale $\mu$ in the parton
distribution functions, subprocess cross sections, and the fragmentation
functions to be the same.  We will later vary $\mu$ between $0.5 \mu_R$ and
$2\mu_R$, where $\mu_R$ is given in Eq.~(\ref{scale}), to study the
dependence on the choice of scale.
For the strong coupling constant $\alpha_s(\mu)$ entered in the subprocess
cross sections, we employ the following simple expression
\begin{equation}
\alpha_s (\mu) = \frac{\alpha_s(M_Z)}{1+ \;\frac{33-2n_f}{6\pi}\; \alpha_s(M_Z)
\ln(\frac{\mu}{M_Z}) }\;,
\end{equation}
where $n_f$ is the number of active flavors at the scale $\mu$ and
$\alpha_s(M_Z)=0.118$.
In order to simulate the detector coverage at the Tevatron,
we impose the following acceptance cuts on the transverse momentum and
rapidity of the $(\bar b c)$ state $H$:
\begin{equation}
p_T(H) > 6\;{\rm GeV} \qquad {\rm and} \qquad |y(H)| < 1 \;.
\end{equation}

The numerical results of the $p_T$ spectra for the $(\bar b c)$ state $H$
with various spin-orbital
quantum numbers are shown in Fig.~\ref{fig1} and Fig.~\ref{fig2}
for the cases of principal quantum number $n=1$ and $n=2$, respectively.
The integrated cross sections versus $p^{\rm min}_{T}(H)$ are also shown
in Figs.~\ref{fig3} and \ref{fig4} for the cases of $n=1$ and $n=2$,
respectively.

Now we can predict the inclusive production rate of the $B_c$ meson.
As the annihilation channel is suppressed relative to the
electromagnetic/hadronic transitions, all the $(\bar b c)$
excited states below  the
$BD$ threshold will decay eventually into the
ground state $B_c$ via emission of photons or pions.
Since the energies of these emitted photons and/or pions are limited by
the small mass differences between the initial and final $(\bar b c)$ states,
the transverse momenta of the $(\bar bc)$ mesons will not be altered
appreciably during the cascades. Therefore, we can simply
add up the $p_T$ spectra (Fig.~\ref{fig1} and Fig.~\ref{fig2}) of
all the states to represent the $p_T$ spectrum of the inclusive
$B_c$ production.
Similarly, we can add up the integrated cross sections (Fig.~\ref{fig3}
and Fig.~\ref{fig4}) to represent the integrated  cross section for the
inclusive  $B_c$ production.
Thus, we can obtain the inclusive production rate of $B_c$ as a function
of $p_T^{\rm min}(B_c)$.    Table~\ref{table2}
gives the inclusive  cross sections for the $B_c$ meson
at the Tevatron as a function of $p_T^{\rm min}(B_c)$,
including all the contributions from the $n=1$ and $n=2$ S-wave
and P-wave states.   These cross sections
should almost represent the inclusive production of $B_c$ by fragmentation,
because the contributions from the D-wave states are expected to be minuscule.
The results for $\mu=\mu_R/2,\,\mu_R$, and
$2\mu_R$ are also shown.   The variation of the integrated cross sections
with the scale $\mu$ is always within
a factor of two, and only about 20\% for $p_T>10$ ~GeV.
We will discuss more about the dependence of scale in the next section.

At the end of Run Ib at the Tevatron, the total accumulated luminosities
can be up to 100--150 pb$^{-1}$ or more.  With $p_T>6$~GeV, there are about
$5 \times 10^5$ $B_c^+$ mesons.
The lifetime of the $B_c$ meson has been estimated to be of order 1--2
picosecond \cite{eitqui}, which is long enough to leave a displaced
vertex in a silicon vertex detector.  Besides, $B_c$ decays into $J/\psi+X$
very often, where $X$ can be a $\pi^+$, $\rho^+$, or $\ell^+ \nu_l$, and
$J/\psi$ can be detected easily through its leptonic decay modes.
The inclusive branching ratio of $B_c\to J/\psi +X$
is about  10\% \cite{mangano}.
When $X$ is $e^+\nu_e$ or $\mu^+\nu_\mu$, we will obtain the striking
signature of three-charged leptons coming off from a common
secondary vertex. The combined branching ratio of
$B_c\to J/\psi \ell^+\nu_\ell \to \ell'^+ \ell'^- \ell^+ \nu_\ell\;
(\ell,\ell'=e,\mu)$ is about 0.2\%.  This implies that there will be of order
$10^3$ such distinct events for 100 pb$^{-1}$ luminosity at the Tevatron.
Even taking into account the imperfect detection efficiencies,
there should be enough events for confirmation.
However, this mode does not afford  the full reconstruction of the $B_c$.
If $X$ is some hadronic states, {\it e.g.}, pions, the events can be
fully reconstructed and the $B_c$ meson mass can be measured.
The process $B_c\to J/\psi+ \pi^+ \to \ell^+\ell^- \pi^+$ is  likely to be
the discovery mode for $B_c$. Its combined branching ratio is
about 0.03\%, which implies about 300 such distinct
events  at the Tevatron with a luminosity of  100 pb$^{-1}$.

After the next fixed target runs at the Tevatron, the Main Injector will be
installed in 1996--1997 according to the present plan \cite{talk}.
The Main Injector will give a significant boost in the luminosity while
the center-of-mass energy stays the same.  The upgraded luminosity
is estimated to be about ten times larger than its present value.
This enables Run II to accumulate a total luminosity of 1--2 fb$^{-1}$,
which implies that about $10^7 - 10^8$ $B_c$ mesons will be produced.
With the Main Injector installed, it might be
possible to produce the D-wave $(\bar b c)$ states with sizable rates.

Tevatron will continue running until the next generation of hadronic
colliders, {\it e.g.}, the LHC.  The present design of the LHC is
at a center-of-mass energy of 14 TeV and the yearly luminosity is of
order 100 fb$^{-1}$.
In Table~\ref{table3}, we show the inclusive cross sections for
the $B_c$ meson at the LHC as a function of $p_T^{\rm min}(B_c)$
including the contributions from $n=1$ and $n=2$ S-wave and P-wave states.
With the assumed 100 fb$^{-1}$ luminosity there are about $3\times 10^9$
$B_c$ mesons with $p_T> 10 $~GeV and $|y(B_c)|<2.5$.
With such a high luminosity at LHC,
one expects sizable numbers of the various D-wave $(\bar b c)$ states to be
produced as well.
The LHC will then be a copious source of $(\bar b c)$ mesons
such that their properties {\it e.g.}, spectroscopy and decays,
can be thoroughly studied. In addition, the mixing and CP violation studies
are possible. For example, one can use $B_c$ to tag the flavor
of $B_s$ in the decay $B_c^+ \to B_s^0 \ell^+ \nu_l$ for the
studies of $B_s^0 - \bar{B_s^0}$ mixing.
Also, CP violations in the $B_c$ system can be studied by looking at the
difference in the partial decay widths of $B_c^+ \to X$ and $B_c^- \to \bar X$.

\section{Discussions and conclusions}

First, we briefly discuss the various sources of uncertainties in our
calculation. One uncertainty comes from the use of the naive
Altarelli-Parisi evolution  equations.
As pointed out in Refs.~\cite{BDFM,BCFY},
the naive Altarelli-Parisi equations in Eqs.~(\ref{Db})--(\ref{Dg}) do not
respect the phase space constraints.  Inhomogeneous evolution equations were
then advocated to remedy for these problems.  The major effect is to correct
the unphysical blow-up of the evolved gluon fragmentation functions when
$z$ gets too close to 0.  The corrected gluon fragmentation functions,
instead of having the unphysical blow-up at $z=0$, turn to
zero smoothly below a certain threshold value of $z$.
Despite the dramatic changes of the evolved gluon fragmentation functions
for small $z$ values, this effect will not show up easily in our calculation
because the small $z$ region is very likely excluded by the transverse
momentum cut imposed on the $(\bar bc)$ mesons.
Therefore, in this paper we keep on using
the homogeneous Altarelli-Parisi equations in
Eqs.~(\ref{Db})--(\ref{Dg}) to evolve our fragmentation functions.

A second source of uncertainty is due to the choice of the factorization
scale $\mu$. We show in Fig.~\ref{fig5}
the dependence of the differential cross sections on the
factorization scale $\mu$ by plotting the results for the various
choices of $\mu =\mu_R/2, \mu_R$, and $2\mu_R$.
For clarity we only show the curves for the
$1\,^1S_0$ and $1\,^3P_0$ states in Fig.~\ref{fig5}.
The behaviors for other states are similar.  Note that
in the $\mu=\mu_R/2,\,\mu_R$, and $2\mu_R$ curves, the running scales
used in the strong coupling constant $\alpha_s$, which entered
in $d \hat \sigma$'s
and in the parton distribution functions $f_i(x)$'s are equal to $\mu$,
while the running scale used in the fragmentation functions
is set to be ${\rm max}$ $(\mu,\,\mu_0)$, where $\mu_0$
is the prescribed initial scale for the fragmentation
functions.  Figure~\ref{fig5} shows
that the change in the factorization scale $\mu$ gives different
results for the differential cross sections.   The $\mu=2\mu_R$
curves show that the differential
cross sections increase (decrease) only slightly at the low (high)
$p_T$ region. Although the $\mu=\mu_R/2$ curves demonstrate larger changes in
the differential cross sections, the variations are always within
a factor of two. The integrated cross sections at various
$p^{\rm min}_{T}(B_c)$, as already shown in
Table~\ref{table2}, also indicate lesser sensitivity in the scale
as one increases $p^{\rm min}_{T}(B_c)$.
The variations of differential
cross sections with the scale $\mu$ demonstrate  the
effects of the next-to-leading order (NLO) corrections.  Only if all the NLO
corrections are calculated for  each piece contained in Eq.~(\ref{*}),
namely the parton distribution functions, the parton-parton hard scattering
cross sections, and the fragmentation functions, can these variations be
reduced substantially.  Since the perturbative QCD fragmentation functions
are only calculated to leading order,  the NLO calculations of
the perturbative QCD fragmentation functions can provide an improvement to our
calculation.   However, due to the rather weak dependence on the scale
in our leading order calculation, our results should be rather stable
under higher order perturbative corrections.

Other uncertainties come from the input parameters to the boundary
conditions of the fragmentation functions.  These are the heavy
quark masses $m_b$ and $m_c$, and the nonperturbative parameters
describing the bound states. Slight
changes in $m_c$ and $m_b$ could possibly lead to appreciable changes
in the fragmentation functions, as indicated by the $m^3$ and $m^5$
dependence, respectively in the denominators of the S-wave
and P-wave fragmentation functions. (Note that
$H_1/m \approx 9 \vert R^\prime(0) \vert^2/(32\pi m^5)$.)
However, in the numerators
the wavefunctions at the origin $\vert R(0) \vert^2$ and
$\vert R^\prime(0) \vert^2$ also
scale like $m^3$ and $m^5$, respectively. Therefore
the dependence on the heavy quark masses
should be  mild in the fragmentation functions.
The color-octet parameters $H_8'$'s are associated with the additional
color-octet contributions and they are not well
determined. However, we do not expect that they will play any significant role
in the present context and only refer our readers
to Ref.\cite{qpwave} where some discussions of these parameters can be found.

There was a controversy in the recent literature
\cite{changchen,slab,bere,mase,klr} concerning
about the importance of parton fragmentation in the
production of the $B_c\;(B_c^*)$ meson at hadronic colliders.
The controversy arises because the Feynman diagrams responsible
for the $\bar b$ antiquark fragmentation form a
 subset of the whole set of Feynman diagrams, which
contribute at the order of $\alpha_s^4$.
Thus, there is a competition between the fragmentation contribution and the
non-fragmentation contribution. Some authors had referred to the
latter contribution as recombination.
So far, five independent groups \cite{changchen,slab,bere,mase,klr}
have presented such a complete ${\cal O}(\alpha_s^4)$ calculation.
Chang {\it et al}. \cite{changchen} and Slabospitsky \cite{slab}
agreed that the fragmentation contribution dominates at the large transverse
momentum region. However,  we could not find in their work the
precise value of $p_T$ at which the fragmentation contribution begins
to dominate. On the other hand, Berezhnoy {\it et al.} \cite{bere}
claimed that fragmentation never dominates for all
kinematical region of $p_T$ and the recombination diagrams
can never be ignored. Independently, Masetti and Sartogo \cite{mase}
have recently obtained results claimed to be consistent
with Berezhnoy {\it et al} \cite{bere}.
There are certainly discrepancies among these calculations.
Most recently,  Kolodziej {\it et al} \cite{klr} performed yet another
independent calculations using two different methods --
the usual trace and the modern helicity amplitude techniques.
Both methods agreed with each other numerically to very high accuracy,
according to the authors.   Kolodziej {\it et al.} \cite{klr}
also did a very thorough
comparison with previous exact calculations by  comparing numerically
their matrix elements with other groups \cite{changchen,slab,bere},
but could not find agreement using identical input parameters.
\footnote{In their most recently revised versions,
both groups of Chang {\it et al.}
(the second entry of Ref.~\cite{changchen}) and
Berezhnoy {\it et al.} \cite{bere}
confirmed the results of Kolodziej {\it et al.} \cite{klr}.
Thanks to an anonymous referee for pointing out the situation.}
Furthermore, Kolodziej {\it et al.} \cite{klr}
did a comparison of the production of
$B_c$ mesons between the exact calculation and the fragmentation
approximation, and found that the fragmentation approach is indeed valid for
$p_T \ge 10$ GeV!
%
%
%
%
%
Because of the conclusions of Ref.~\cite{klr},
we believe that the fragmentation contribution should begin to dominate
the production of $(\bar bc)$ mesons at $p_T \agt 10$ GeV.
Also, we would like to emphasize that the detector coverage and performance
do require a minimum transverse momentum cut and a rapidity cut on the
$B_c$ meson. Thus, the non-fragmentation (recombination) contribution,
which is substantial at low $p_T$, becomes less important as the
lowest $p_T$ range is being excluded.

The main result of this paper is the inclusive cross section
of the ground state $B_c$ at the Tevatron. We have included in this work
all the contributions from the direct $\bar b$ fragmentation and induced gluon
fragmentation, as well as the cascade contributions
{}from all the excited S-wave and P-wave states below the $BD$ threshold.
Previous result presented by one of us in Ref.\cite{cheung} only included the
direct $\bar b$ fragmentation into S-wave states, since the P-wave
fragmentation
functions were not available at that time.  Later it was pointed out in
Ref.\cite{induceglue} that the induced gluon fragmentation contribution is
nonnegligible.  The induced gluon fragmentation contribution
is about 20 -- 30\% (30 -- 40\%) of the direct $\bar b$
fragmentation contribution at the Tevatron (LHC) energies.
The two sets of P-wave states included in this work contribute at the level
of 30\%.  Furthermore, we have employed here
a more updated set of parton distribution functions than those used in
Refs.\cite{cheung,induceglue}.  As a result, the cross sections increase
roughly by 25\%, mainly due to the increase in the gluon luminosities.
When all these aforementioned differences are taken into account,
the net result is about 70 -- 80\% (a factor of two)
larger than the previous result presented in Ref.\cite{cheung}
at the Tevatron (LHC).

To recap, in this paper we have performed calculations of the
transverse momentum spectra and integrated cross sections for the
S- and P-wave $(\bar b c)$ mesons.  We have also predicted the inclusive
production rate for the $B_c$ meson, and found that there should be enough
signature events to confirm the existence of $B_c$ at the Tevatron,
whereas the LHC will be a copious source of $B_c$.
Nevertheless, one should keep in mind that what we have calculated in
this paper represents only the contribution from fragmentation, while there
should also be other non-fragmentation contributions, especially at
the low $p_T$ region ($p_T \alt 10$ GeV).
At least, our results represent a lower bound of the
production rate for $B_c$.  In the near future, the
production of $B_c$  at the Tevatron would provide another
interesting test for perturbative QCD, and we expect a very exciting
experimental program of $B_c$ at the LHC.

\section*{Acknowledgement}
This work was supported in part by the United States Department of
Energy under Grant Numbers DE-FG03-93ER40757
and DE-FG03-91ER40674.

\newpage
\appendix
\section*{Appendix}
\setcounter{equation}{0}
\renewcommand{\theequation}{A.\arabic{equation}}

The Altarelli-Parisi evolution equations were given in
Eqs.~(\ref{Db})--(\ref{Dg}).  For convenience,
we also list all the Altarelli-Parisi kernels in the following,
\begin{eqnarray}
\label{Pbb}
P_{\bar b\to\bar b}(x,\mu) &=&
\frac{4\alpha_s(\mu)}{3\pi} \left( \frac{1+x^2}{1-x} \right )_+ \,, \\
\label{Pbg}
P_{\bar b \to g}(x,\mu) &=& \frac{4\alpha_s(\mu)}{3\pi} \left(
\frac{1+(1-x)^2}{x} \right )_+ \,, \\
\label{Pgb}
P_{g\to\bar b}(x,\mu) &=& \frac{\alpha_s(\mu)}{2\pi} \left( x^2+(1-x)^2 \right)
\,, \\
\label{Pgg}
P_{g \to g}(x,\mu) &=& \frac{6 \alpha_s(\mu)}{\pi} \left(
{x \over (1-x)_+} + {1-x \over x} + x(1-x) + \left(
\frac{11}{12}-\frac{n_f}{18} \right ) \delta (1-x) \right) \, ,
\end{eqnarray}
where $n_f$ is the number of active flavors below the scale $\mu$.

The fragmentation functions $D_{\bar b \to H}(z,\mu_0)$ for
the S-wave and P-wave $(\bar b c)$ states have been obtained
to leading order in strong coupling constant $\alpha_s$
at the heavy quark mass scale $\mu_0 = m_b + 2 m_c$.
We introduce the following notations:
$r = m_c/(m_b+m_c)$, $\bar r = 1-r$, and $m = m_b m_c/(m_b + m_c)$.
The nonperturbative parameters are $R_{nS}(0)$ for the S-wave states,
and $H_1(n)$ and $H_8^\prime(n)$ for the P-wave states.
The mixing angle between the $^1P_1$ and $^3P_1$ states is
denoted by $\cos \theta_{\rm mix}$.
The numerical values of these input parameters can be inferred from
potential model calculations (see for example \cite{eitqui})
and are listed in Table~\ref{table1} for convenience.
The scale of the strong coupling constant entered in
$D_{\bar b \to H}(z,\mu_0)$ is set to be $2 m_c$.

The S-wave fragmentation functions are \cite{bcfrags}
\begin{eqnarray}
\label{DBc}
D_{{\bar b} \rightarrow \bar b c(n^1S_0)}(z,\mu_0)
& = & {2 \alpha_s(2 m_c)^2 |R_{nS}(0)|^2 \over 81 \pi m_c^3}
\; {r z (1-z)^2 \over (1 - \bar r z)^6} \nonumber \\
& \times & \Bigg[ 6 \;-\; 18(1-2r)z
\; + \;(21-74r+68r^2)z^2 \nonumber \\
&& - \, 2\bar r(6-19r+18r^2)z^3 \, + \, 3\bar r^2(1-2r+2r^2)z^4 \Bigg]
\;, \nonumber \\
&& \;
\end{eqnarray}
\vfill
\newpage
\begin{eqnarray}
D_{{\bar b} \rightarrow \bar b c(n^3S_1)}(z,\mu_0)
& =& {2 \alpha_s(2m_c)^2 |R_{nS}(0)|^2 \over 27 \pi m_c^3} \;
{r z (1-z)^2 \over (1 - \bar r z)^6} \nonumber \\
& \times & \Bigg[ 2 \;-\; 2(3-2r)z
\; + \; 3(3-2r+4r^2)z^2  \;-\; 2\bar r(4-r+2r^2)z^3 \nonumber \\
&&	\;\;\; +\; \bar r^2(3-2r+2r^2)z^4 \Bigg] \;.
\label{DBstar} \end{eqnarray}
Polarized S-wave fragmentation functions can also be
found in Ref.~\cite{bc_pol} but they are not used in this work.

The P-wave fragmentation functions consist of two terms,
the color-singlet piece ($H_1$ term)
and the color octet piece ($H_8^\prime$ term) \cite{qpwave}:
\begin{eqnarray}
\label{d1p1total}
D_{\bar b \to \bar b c(n^1P_1)}(z,\mu_0) & = & {H_{1(\bar bc)}(n) \over m}
D^{(1)}_{\bar b \to \bar b c(^1P_1)}(z,\Lambda) +
3 {H'_{8(\bar b c)}(\Lambda) \over m}
D^{(8)}_{\bar b \to  \bar b c(^1S_0)}(z) \; , \\
\label{d3pjtotal}
D_{\bar b \to \bar b c(n^3P_J)}(z,\mu_0) & = & {H_{1(\bar b c)}(n) \over m}
D^{(1)}_{\bar b \to \bar b c(^3P_J)}(z,\Lambda)
+ (2J+1) {H'_{8(\bar b c)}(\Lambda) \over m}
D^{(8)}_{\bar b \to \bar b c(^3S_1)}(z) \; , \nonumber \\
&& \;
\end{eqnarray}
where $\Lambda$ is a factorization scale for the P-wave fragmentation
functions. It separates the reduced heavy quark mass scale $m$ and
the scale $m v$ which characterizes the bound state structures
with $v$ being the typical relative velocity of the heavy quark and
antiquark inside the bound state. In our numerical work,
we will choose $\Lambda \sim m$.

At leading order the color-singlet pieces do not depend on $\Lambda$ and
they are given by
\begin{eqnarray}
D^{(1)}_{\bar b \rightarrow \bar b c(^1P_1)} (z) & = &
{16 \alpha_s^2(2m_c) \over 243}
{r \bar r^3 z (1-z)^2 \over (1 - \bar r z)^8}
\Bigg[ 6 - 6 (4r^2-8r+5) z  \nonumber \\
& + & ( 32r^4-96r^3+250r^2-210r+69) z^2 \nonumber \\
& + & 8 \bar r (4r^4+12r^3-48r^2+37r-12) z^3  \nonumber \\
& + & 2 \bar r^2 (16r^4+161r^2-114r+42) z^4
- 6 \bar r^3 (4r^3+28r^2-15r+7) z^5 \nonumber \\
& + &   \bar r^4 (46r^2-14r+9) z^6 \Bigg] \; ,
\label{dz1p1}
\end{eqnarray}
\vfill
\newpage
\begin{eqnarray}
D^{(1)}_{\bar b \rightarrow \bar b c(^3P_0)} (z) & = &
{16 \alpha_s^2(2m_c) \over 729}
{r \bar r^3 z (1-z)^2 \over  (1 - \bar r z)^8} \Bigg[
6(4r - 1)^2 + 6(4r-1)(20r^2-16r+5) z  \nonumber \\
& + & ( 832r^4-1456r^3+1058r^2-362r+63) z^2 \nonumber \\
& - & 8 \bar r (100r^4-184r^3+118r^2-22r+9) z^3 \nonumber \\
& + & 2 \bar r^2 (416r^4-776r^3+369r^2+42r+24) z^4 \nonumber \\
& - & 2 \bar r^3 (240r^4-516r^3+232r^2+59r+9) z^5 \nonumber \\
& + &   \bar r^4 (96r^4-240r^3+134r^2+34r+3) z^6 \Bigg] \; ,
\label{dz3p0}
\end{eqnarray}
\begin{eqnarray}
D^{(1)}_{\bar b \rightarrow \bar b c(^3P_1)} (z) & = &
{32 \alpha_s^2(2m_c) \over 243}
{r \bar r^3 z (1-z)^2 \over (1 - \bar r z)^8} \Bigg[
6 + 6 (4r-5) z  \nonumber \\
& + & ( 16r^4+64r^2-98r+63) z^2
   + 8\bar r (2r^4+2r^3-13r^2+11r-9) z^3  \nonumber \\
& + & 2 \bar r^2 (8r^4-16r^3+47r^2-18r+24) z^4
   + 2 \bar r^3 (8r^3-24r^2+r-9) z^5 \nonumber \\
& + &   \bar r^4 (12r^2+2r+3) z^6 \Bigg] \; ,
\label{dz3p1}
\end{eqnarray}
and
\begin{eqnarray}
D^{(1)}_{\bar b \rightarrow \bar b c(^3P_2)} (z) & = &
{64 \alpha_s^2(2m_c) \over 729}
{r \bar r^5 z (1-z)^2 \over  (1 - \bar r z)^8} \Bigg[
12  + 12 (2r-5) z  \nonumber \\
& + &  ( 92r^2-76r+135) z^2
   + 4 (10r^3-54r^2+31r-45) z^3 \nonumber \\
& + & 2 (46r^4-16r^3+123r^2-78r+75) z^4 \nonumber \\
& - & 4 \bar r (6r^4+9r^3+40r^2-13r+18) z^5 \nonumber \\
& + &   \bar r^2 (12r^4-12r^3+55r^2-10r+15)  z^6 \Bigg] \; .
\label{dz3p2}
\end{eqnarray}

The $^1P_1$ and $^3P_1$ states are mixed in general,  giving
rise to the following
physical mass eigenstates $|1^{+'}\rangle$ and
$|1^{+}\rangle$,
\begin{eqnarray}
\label{mixing1}
|1^{+'}\rangle  & = & \cos \theta_{\rm mix} |^1P_1 \rangle
		\; - \; \sin \theta_{\rm mix} |^3P_1 \rangle \; , \\
|1^+ \rangle & = &  \sin \theta_{\rm mix} |^1P_1 \rangle
		\; + \; \cos \theta_{\rm mix} |^3P_1 \rangle
\; .
\label{mixing2}
\end{eqnarray}
\vfill
\newpage

\noindent
Thus, in general, we have
\begin{eqnarray}
D^{(1)}_{\bar b \rightarrow \bar b c(1^{+'})} (z) & = &
\cos^2\theta_{\rm mix} D^{(1)}_{\bar b \rightarrow \bar b c(^1P_1)} (z) +
\sin^2\theta_{\rm mix}  D^{(1)}_{\bar b \rightarrow  \bar b c(^3P_1)} (z)
\nonumber
\\
&& -\sin\theta_{\rm mix} \cos\theta_{\rm mix} D^{(1)}_{\rm mix} (z) \; , \\
D^{(1)}_{\bar b \rightarrow \bar b c(1^+)} (z) & = &
\sin^2\theta_{\rm mix} D^{(1)}_{\bar b \rightarrow \bar bc(^1P_1)} (z) +
\cos^2\theta_{\rm mix}  D^{(1)}_{\bar b \rightarrow  \bar b c(^3P_1)} (z)
\nonumber
\\
&& +\sin\theta_{\rm mix} \cos\theta_{\rm mix} D^{(1)}_{\rm mix} (z) \; ,
\label{dzmix}
\end{eqnarray}
with
\begin{eqnarray}
D^{(1)}_{\rm mix} (z) & = & - {32 \sqrt 2 \alpha_s^2(2m_c) \over 243}
		{r \bar r^3 z (1-z)^2 \over  (1 - \bar r z)^6}
\Bigg[   6 - 6(2r^2-4r+3) z  \nonumber \\
& + & (24r^3+52r^2-52r+21) z^2
  + 2 \bar r (14r^3-6r^2+15r-6) z^3 \nonumber \\
& - &  \bar r^2 (2r^2+8r-3) z^4 \Bigg] \; .
\label{dzmix1}
\end{eqnarray}

Finally, the color-octet pieces are given by
\begin{eqnarray}
D^{(8)}_{\bar b \to \bar b c(^1S_0)}(z) & = & {\alpha_s^2(2m_c) \over 162}
{r \bar r^3 z(1-z)^2 \over (1-\bar r z)^6}
\Bigg[  6 - 18(1 - 2r)z + (21 - 74r + 68r^2)z^2 \nonumber \\
&&\;\;\;\;\;\;\; \;\;\;
- \; 2\bar r(6 - 19r + 18r^2)z^3 + 3\bar r^2(1 - 2r + 2r^2)z^4 \Bigg] \; ,
\label{d81s0}
\end{eqnarray}
and
\begin{eqnarray}
D^{(8)}_{\bar b \to \bar b c(^3S_1)}(z) & = &
{\alpha_s^2(2m_c) \over 162}
{r \bar r^3 z(1-z)^2 \over (1-\bar r z)^6}
\Bigg[  2 - 2(3-2r)z + 3(3 - 2r + 4r^2)z^2 \nonumber \\
&&\;\;\;\;\;\;\; \;\;\;
- \; 2 \bar r(4 - r + 2r^2)z^3 + \bar r^2(3 - 2r + 2r^2)z^4 \Bigg] \, .
\label{d83s1}
\end{eqnarray}
%


\newpage
\begin{table}[h]
\caption[]{\small Input parameters to the perturbative QCD fragmentation
functions for $n=1$ and $n=2$.
\label{table1}
}
\medskip
\centering
\begin{tabular}{c@{\extracolsep{0.5in}}cc}
\hline
\hline
       &   $n=1$    &    $n=2$ \\
\hline
$m_b$    &  4.9 GeV  &   4.9 GeV \\
$m_c$    &  1.5 GeV  &   1.5 GeV \\
$R_{nS}(0)$   &  1.28 GeV$^{3/2}$  &   0.99 GeV$^{3/2}$ \\
$H_1$      &  10 MeV   &  14 MeV \\
$H_8'(m)$     &  1.3 MeV \cite{qpwave}  &  1.8 MeV \cite{qpwave}\\
$\cos\theta_{\rm mix}$  & 0.999 &  0.957 \\
\hline
\hline
\end{tabular}
\end{table}

\begin{table}[h]
\caption[]{\small The  inclusive production cross sections for
the $B_c$ meson at the Tevatron including the contributions from all the
S-wave and P-wave states below the $BD$ threshold as a function
of $p_{T}^{\rm min}(B_c)$.
The acceptance cuts are $p_T(B_c)>6$ GeV and $|y(B_c)|<1$.
\label{table2}
}
\bigskip
\bigskip
\centering
\begin{tabular}{c@{\extracolsep{0.5in}}ccc}
\hline
\hline
 $p_{T}^{\rm min}$ (GeV)  &  \multicolumn{3}{c}{$\sigma$ (nb)} \\
\hline
     &\underline{$\mu = \frac{1}{2}\mu_R$} & \underline{$ \mu=\mu_R$}
 & \underline{$\mu=2 \mu_R$} \\
 6         &  2.81  &  5.43  &  6.93 \\
10         &  0.87  &  1.16  &  1.22 \\
15         &  0.26  &  0.29  &  0.26 \\
20         &  0.098 &  0.097 &  0.083\\
30         &  0.021 &  0.018 &  0.014 \\
\hline
\end{tabular}
\end{table}


\newpage
\begin{table}[t]
\caption[]{The inclusive production cross sections
for the $B_c$ meson at the LHC
including the contributions from all the S-wave
and P-wave states  below the $BD$ threshold as a function of
$p_T^{\rm min}(B_c)$.
The acceptance cuts are $p_T(B_c)>10$ GeV and $|y(B_c)|<2.5$.
\label{table3}
}
\bigskip
\bigskip
\centering
\begin{tabular}{c@{\extracolsep{0.6in}}c}
\hline
\hline
 $p_T^{\rm min}$ (GeV)  &  $\sigma$ (nb), $\mu=\mu_R$ \\
\hline
10         & 33.8\\
20         &  4.4 \\
30         & 1.1\\
40         & 0.41 \\
\hline
\end{tabular}
\end{table}

\bigskip
\bigskip

\begin{center}
\section*{Figure Captions}
\end{center}

\begin{enumerate}

\item
\label{fig1}
The differential cross section $d\sigma/p_T$ versus $p_T$ of the
$(\bar b c)$ meson ($H$) in various spin-orbital states
with $n=1$ at the Tevatron. The acceptance cuts are $p_T(H)>6$ GeV
and $|y(H)|<1$.

\item
\label{fig2}
Same as Fig.1 for $n=2$.

\item
\label{fig3}
The integrated cross section $\sigma(p_T > p_T^{\rm min})$ versus
the minimum $p_T^{\rm min}$ cut on the $(\bar b c)$ meson ($H$) in various
spin-orbital states with $n=1$ at the Tevatron.  The acceptance cuts are
the same as Fig.1.

\item
\label{fig4}
Same as Fig.3 for $n=2$.

\item
\label{fig5}
The comparison of the differential cross sections $d\sigma/dp_T$
at different factorization scales for the $1\;^1S_0$ and $1\;^3P_0$
states at the Tevatron.
The acceptance cuts are the same as Fig.1.

\end{enumerate}


\newpage
\begin{thebibliography}{99}
%
\bibitem{CDF}F. Abe {\it et al.} (CDF Coll.), Phys. Rev. Lett. {\bf 69}, 3704
(1992).
%
\bibitem{stirling}E. W. N. Glover, A. D. Martin, and W. J. Stirling, Z. Phys.
{\bf C38}, 473 (1988).
%
\bibitem{gswave}E. Braaten and T. C. Yuan, Phys. Rev. Lett. {\bf 71}, 1673
(1993).
%
\bibitem{charm}E. Braaten, K. Cheung, and T. C. Yuan, Phys. Rev. {\bf D48},
4230 (1993).
%
\bibitem{bcfrags}E. Braaten, K. Cheung, and T. C. Yuan, Phys. Rev. {\bf D48},
R5049 (1993).
%
\bibitem{gpwave}E. Braaten and T. C. Yuan, Phys. Rev. {\bf D50}, 3176 (1994).
%
\bibitem{qpwave}T. C. Yuan, Phys. Rev. {\bf D50}, 5664 (1994).
%
\bibitem{bcfrags_others}
C.-H. Chang and Y.-Q. Chen, Phys. Lett. {\bf B284}, 127 (1992);
Phys. Rev. {\bf D46}, 3845 (1992), {\bf D50}, 6013 (1994);
Y.-Q. Chen, Phys. Rev. {\bf D48}, 5158 (1993), {\bf D50}, 6013 (1994).
%
\bibitem{BDFM}E. Braaten, M. Doncheski, S. Fleming, and M. Mangano,
Phys. Lett. {\bf B333}, 548 (1994).
%
\bibitem{CG}M. Cacciari and M. Greco, Phys. Rev. Lett. {\bf 73}, 1586 (1994).
%
\bibitem{RS}D. P. Roy and K. Sridhar, Phys. Lett. {\bf B339}, 141 (1994).
%
\bibitem{CWT}P. Cho, M. Wise, and S. Trivedi, Phys. Rev. {\bf D51},
2039 (1995).
%
\bibitem{close}F. E. Close, Phys. Lett. {\bf B342}, 369 (1995).
%
\bibitem{chowise}P. Cho and M. Wise, Phys. Lett. {\bf B346}, 129 (1995).
%
\bibitem{psipi}E. Braaten and S. Fleming, Phys. Rev. Lett. {\bf 74}, 3327
(1995).
%
\bibitem{psipi2}M. Cacciari, M. Greco, M. L. Mangano, and A. Petrelli,
CERN preprint CERN-TH/95-129, hep-ph/9505379 (May 1995).
%
\bibitem{papadimi}Talk given by V. Papadimitriou (CDF Coll.),
FERMILAB-CONF-94-221-E, contribution to DPF'94 meeting at Albuquerque,
NM (August 1994).
%
\bibitem{eitqui}E. Eichten and C. Quigg, Phys. Rev. {\bf D49},
5845 (1994), and references therein.
%
\bibitem{baganetal}
E. Bagan, H. G. Dosch, P. Gosdzinsky, S. Narison, and J.-M. Richard,
Z. Phys. {\bf C64}, 57 (1994).
%
\bibitem{kiselevetal}
V. V. Kiselev, A. K. Likhoded, and A. V. Tkabladze, Serpukhov preprint
IFVE-94-51, hep-ph/9406339 (April 1994);
S. S. Gershtein, V. V. Kiselev, A. K. Likhoded, and A. V. Tkabladze,
Serpukhov preprint IFVE-94-81, hep-ph/9504319 (April 1995).
%
\bibitem{PDG}Particle Data Group, Phys. Rev. {\bf D50}, 1323 (1994).
%
\bibitem{bc_ee}
L. Claveli, Phys. Rev. {\bf D26}, 1610 (1982);
C.-H. Chang and Y.-Q. Chen, Phys. Rev. {\bf D46}, 3845 (1993);
Phys. Lett. {\bf B284}, 127 (1992);
V. V. Kiselev, A. K. Likhoded, and M. V. Shevlyagin, Z. Phys. {\bf C63}, 77
(1994).
%
\bibitem{changchen}
C.-H. Chang and Y.-Q. Chen, Phys. Rev. {\bf D48}, 4086 (1993);
C.-H. Chang, Y.-Q. Chen, G.-P. Han, and H.-T. Jiang, Beijing preprint
AS-ITP-94-24, hep-ph/9408242 and revised version (June 1995).
%
\bibitem{slab}S. R. Slabospitsky, Serpukhov preprint IFVE-94-53,
hep-ph/9404346 (April 1994).
%
\bibitem{bere}A. V. Berezhnoy, A. K. Likhoded, and M. V. Shevlyagin,
Serpukhov preprint IFVE-94-48, hep-ph/9408284 (March 1994);
A. V. Berezhnoy, A. K. Likhoded, and O. P. Yushchenko, Serpukhov preprint
IFVE-95-59, hep-ph/9504302 (April 1995) and revised version.
%
\bibitem{mase}M. Masetti and F. Sartogo, Roma preprint 1099-1995,
hep-ph/9503491 (April 1995).
%
\bibitem{klr}K. Kolodziej, A. Leike, and R. Ruckl, Max Planck Institute
preprint, MPI-PhT/95-36, hep-ph/9505298 (May 1995).
%
\bibitem{cheung}K. Cheung, Phys. Rev. Lett. {\bf 71}, 3413 (1993).
%
\bibitem{induceglue}K. Cheung and T. C. Yuan, Phys. Lett. {\bf B325}, 481
(1994).
%
\bibitem{dwave}K. Cheung and T. C. Yuan, U. T. Austin - U. C. Davis preprint
CPP-95-13/UCD-95-24, hep-ph/9510208 (October 1995).
%
\bibitem{private}F. Abe {\it et al.} (CDF Coll.), FERMILAB-CONF-95-202-E,
paper submitted to Lepton-Photon 95 Conference, Beijing P.R. China, August
1995.
%
\bibitem{cteq}CTEQ Collaboration, J. Botts {\it et al.}, Phys. Lett. {\bf
B304}, 159 (1993).
%
\bibitem{bbl}
G. T. Bodwin, E. Braaten, and G. P. Lepage, Phys. Rev. {\bf D51}, 1125 (1995).
%
\bibitem{mangano}
A. K. Likhoded, S. R. Slabospitsky, M. Mangano, and G. Nardulli,
Nucl. Instrum. Methods {\bf A333}, 209 (1993), and references therein.
%
\bibitem{talk}Talk given by G. Jackson at the
Beyond the Standard Model IV Conference, Tahoe City, CA
(December 13-18, 1994).
%
\bibitem{BCFY}E. Braaten, K. Cheung, S. Fleming, and T. C. Yuan,
Phys. Rev. {\bf D51}, 4819 (1995).
%
\bibitem{bc_pol}K. Cheung and T. C. Yuan, Phys. Rev. {\bf D50}, 3181 (1994).
%
\end{thebibliography}
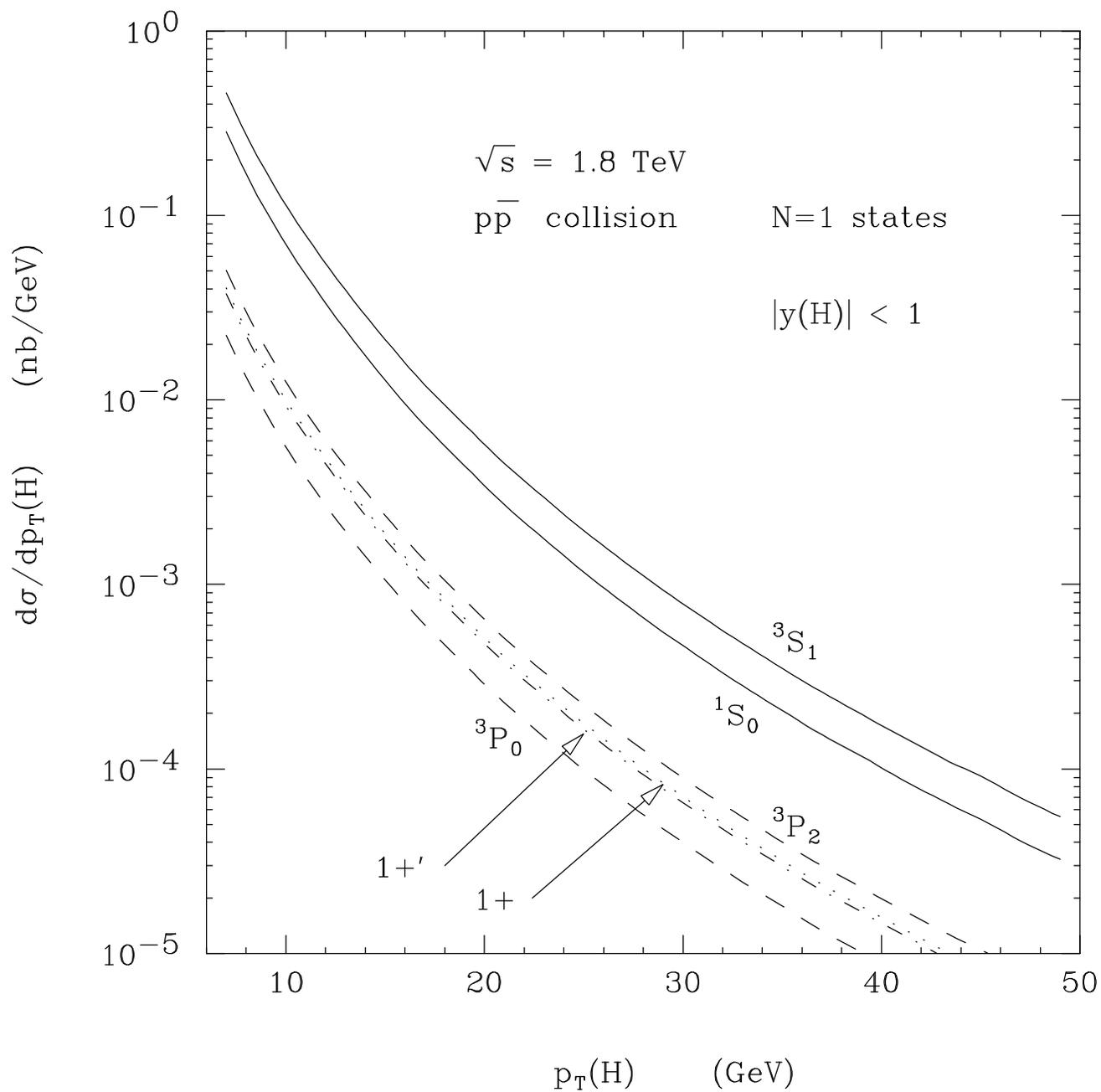
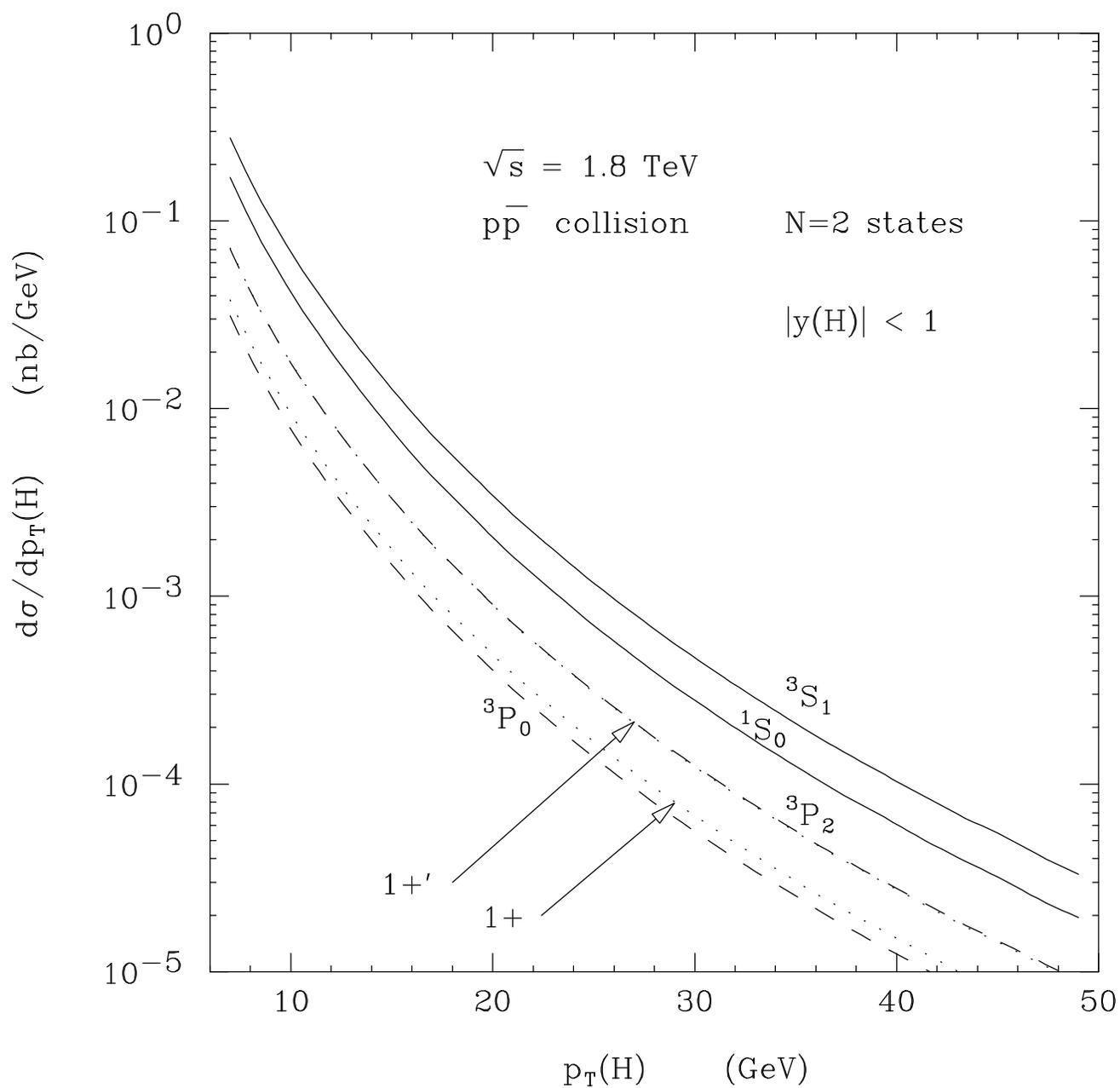
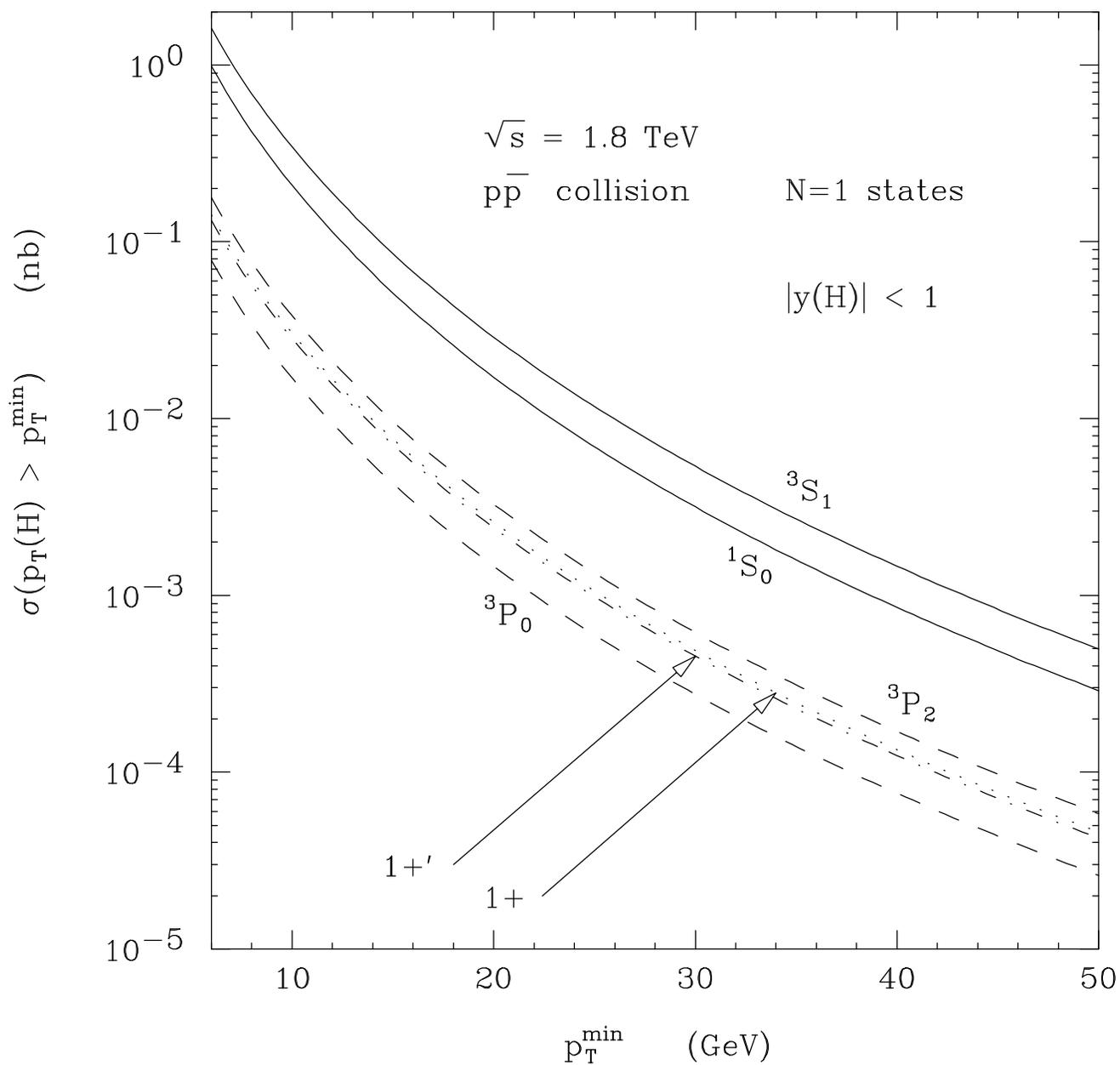
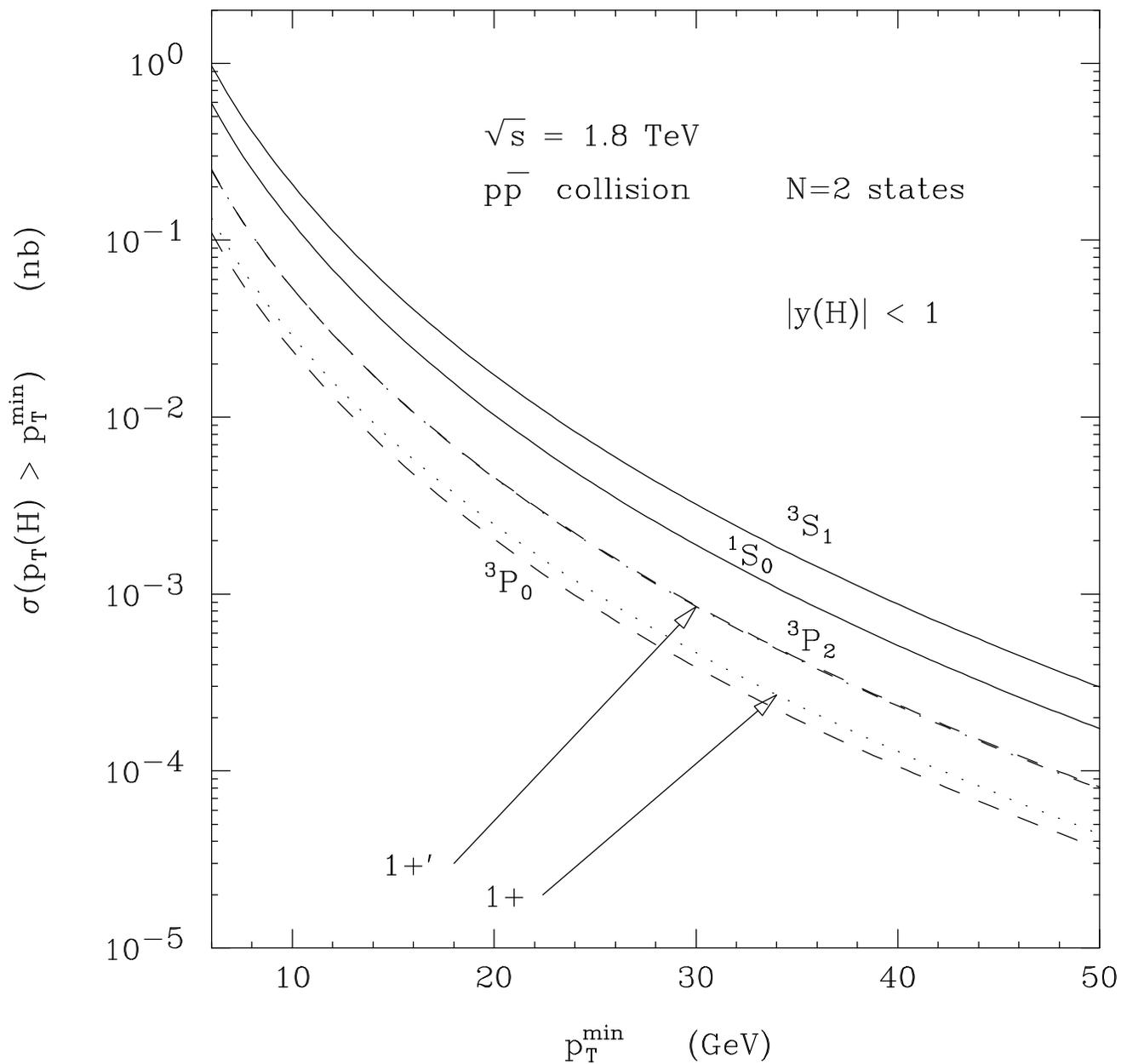
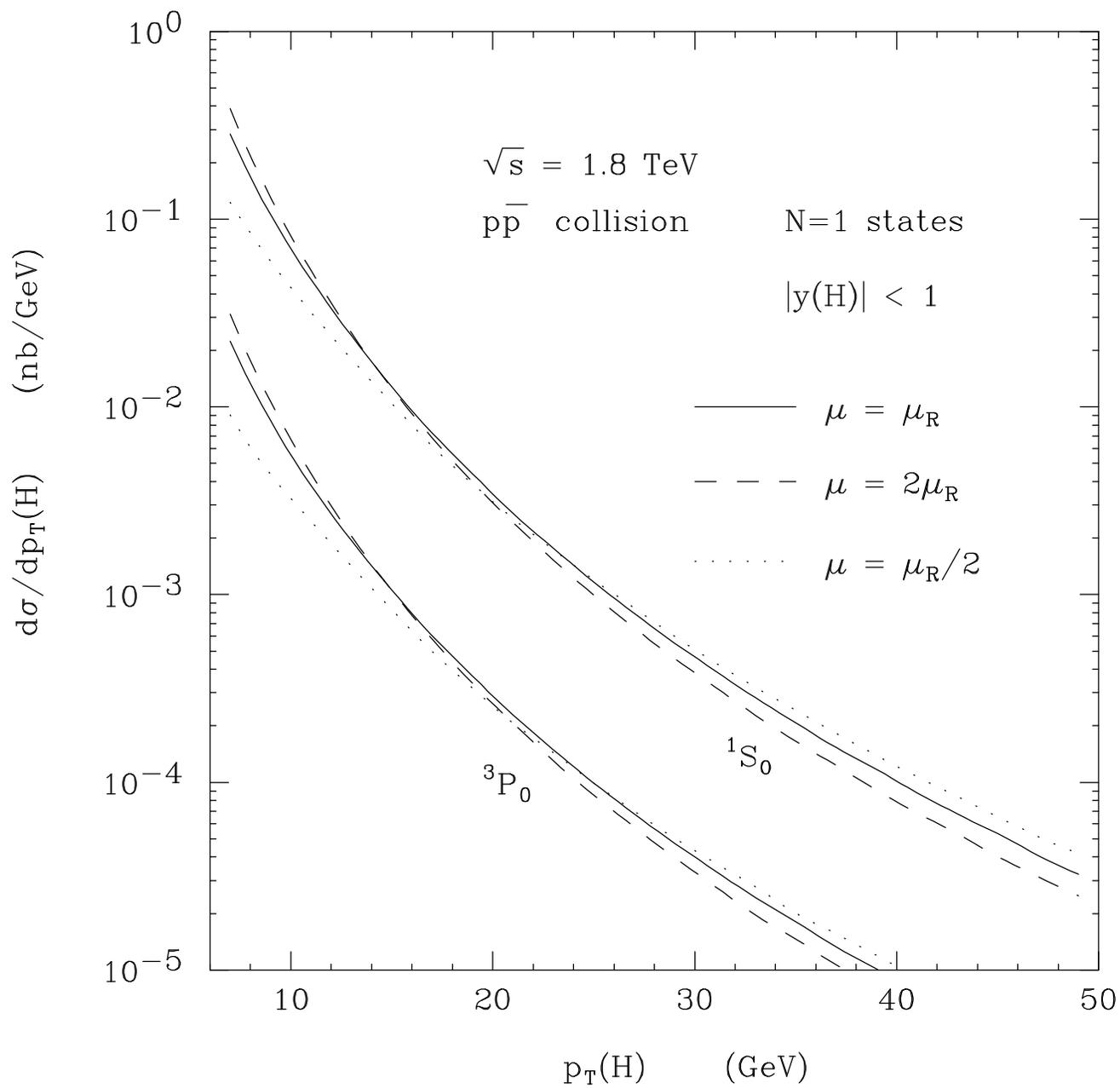
\end{document}